\ifdefined\ANON
\documentclass{Interspeech}
\else
\documentclass[cameraready]{Interspeech}
\fi

\usepackage{booktabs}
\usepackage{graphicx}
\usepackage{iftex}
\usepackage{newtxmath}
\ifXeTeX
\usepackage[no-math]{fontspec}
\fi

\title{VoxLock: Target-Speaker Extraction for Voice Agents Without an Enrollment Step}
\author[affiliation={1}]{Arbaz}{Siddiqui}
\address{$^1$Independent researcher}
\email{arbaz00@gmail.com}


\begin{document}

\maketitle

\begin{abstract}
A voice agent's speech recognizer transcribes everyone near the microphone, including people other than the
caller. I fine-tuned a public causal target-speaker extraction model (REAL-TSE, 27M parameters) on one GPU to work
from a voice sample cut from the call itself, and released it as VoxLock. I score it by what a recognizer writes on
real recordings. On Krisp's 265-recording benchmark it reaches 4.95\% WER against 23.21\% unprocessed and lets 9
background words into the transcript over 85 minutes of background-only speech. Its WER is not significantly
different from Krisp's best filters and is significantly lower than the other commercial and open systems I tested.
On 400 AMI meeting windows built and frozen before scoring, it cuts WER from 42.95\% to 10.63\%. The 8-bit model
runs a live call on 77\% of one CPU thread with 24\,ms of algorithmic latency. Model, code and per-file scores are
open.
\end{abstract}

\section{Introduction}

Voice agents hear whoever is near the microphone: a coworker at the next desk, a television, a second person in
the car. What they say ends up in the transcript, and the agent answers it. Noise suppressors do not help because
they are built to keep speech. Krisp, ai-coustics and Sanas sell ``voice isolation'' filters for this, but they are
closed and licensed per minute or per device.

Target-speaker extraction (TSE) is the research version of the same problem: given a mixture and a short enrollment
recording of one speaker, output only that speaker~\cite{zmolikova2019speakerbeam,wang2019voicefilter,
zmolikova2023overview}. Two things keep it out of voice agents. An agent has no enrollment recording, only the call.
And TSE papers usually report signal metrics on simulated mixtures, while an agent cares about what the recognizer
writes on real audio.

VoxLock is my attempt to close that gap with one GPU. This paper covers:
\begin{itemize}
  \item taking the voice sample from the call itself, so the agent passes in one audio stream;
  \item fine-tuning a public causal TSE baseline~\cite{wang2026realtse} on one NVIDIA L4;
  \item scoring recognizer output against seven models from three commercial vendors and five open models, on
        Krisp's benchmark, 134 AMI windows, and 400 AMI windows frozen before scoring;
  \item releasing the runtime (VoxLock 1.2.1), which runs a live call on one CPU thread.
\end{itemize}

\section{Related work}

SpeakerBeam~\cite{zmolikova2019speakerbeam} and VoiceFilter~\cite{wang2019voicefilter} condition a separator on an
embedding of an enrollment utterance. VoiceFilter-Lite~\cite{wang2020voicefilterlite} made this streaming and
on-device for recognition. The personalized track of the DNS challenges~\cite{dubey2024dns} uses a pre-recorded
enrollment clip. Lin et al.~\cite{lin2023focus} take the speaker cue from the mixture itself instead, using
distance to find the near talker. Band-split RNNs (BSRNN)~\cite{luo2023bsrnn} were extended to personalized
enhancement~\cite{yu2022bsrnnse} and are the TSE backbone in WeSep~\cite{wang2024wesep}. REAL-T~\cite{li2025realt}
and the REAL-TSE challenge~\cite{wang2026realtse} moved evaluation to real conversations and released causal BSRNN
baselines; VoxLock starts from one. Separation models such as TIGER~\cite{xu2025tiger} and SAM
Audio~\cite{shi2025samaudio} output several sources without knowing which is the caller, and noise suppressors such
as DeepFilterNet~\cite{schroeter2023dfn3} keep all speech. I did not find published work comparing open TSE with
commercial voice-isolation products on real recordings by downstream recognition.

\section{Method}

\subsection{Model}
The starting point is the causal \texttt{tfmap\_context} BSRNN from the REAL-TSE baseline in
WeSep~\cite{wang2024wesep,wang2026realtse}: 16\,kHz audio, 512-sample STFT, 128-sample hop, six band-split blocks
and unidirectional sequence modeling, 27M parameters. Its author confirmed that the parameters he trained are
released under Apache-2.0. The voice sample enters twice. First, each frame of the call's spectrum attends over the
frames of the sample's spectrum, giving an extra input channel. Second, the separator's band features attend over
frame-level ECAPA-TDNN~\cite{desplanques2020ecapa} features of the sample, and the result multiplies into each band.
Algorithmic latency is 24\,ms, the 512-sample window minus the 128-sample hop.

\subsection{Voice sample from the call}
The model takes exactly 3\,s of the caller. For a recording I use the loudest 3\,s in the first 15\,s, so a
whole-file run looks up to 15\,s ahead for its sample. For a live stream, audio passes through unchanged until 3\,s
of speech have been heard, counting 10\,ms frames at least 10\,dB above the quietest frame of the last 2\,s; those
become the sample and isolation starts. Both rules assume the caller is the loudest, earliest talker, which holds for
a headset or a phone held by the caller.

\subsection{Training}
Every stage minimizes negative SI-SDR~\cite{leroux2019sisdr} with AdamW (weight decay 0.01, gradient clip 5), trains
all weights including the ECAPA-TDNN branch, and starts from the previous one (Table~\ref{tab:stages}). The learning
rate is $10^{-4}$ in S1 (cosine) and S2, decaying to $10^{-5}$ over the last 1,000 steps of S2, and
$3\times10^{-5}$ in S3 and S4.

\textbf{S1, simulated calls} (4,000 steps, 6\,s examples). A near-field LibriSpeech~\cite{panayotov2015librispeech}
caller is mixed with one or two background talkers, usually 0--15\,dB quieter, in half the examples as close as
the caller and otherwise 1.5--5\,m away in simulated rooms; 25\% of examples go through a telephone channel. The
voice sample is the first 3\,s of the mixture or another utterance of the caller, clean or noisy.

\textbf{S2, real crosstalk} (4,000 steps). Each step adds a batch of AMI headset audio~\cite{carletta2005ami} from
40 meetings to the simulated batch. A headset records its wearer and, more quietly, everyone else. For the first
1,000 steps the real batch is 4\,s windows where the wearer talks alone, an exact target, plus the same headset's
crosstalk from moments when only others talk, at 0--12\,dB above its natural level. For the remaining 3,000 steps it
is 8\,s of that crosstalk alone with a 3--4\,s caller clip pasted in, the target silent outside the clip. No meeting
from the evaluation sets or from REAL-T is used.

\textbf{S3, long calls} (800 steps, 30\,s examples). Trained on 6\,s windows, the model let a caller who talks
without pause fade by 20\,dB within 40\,s. S3 trains on 30\,s examples, most with a continuous caller.

\textbf{S4, distance and room} (400 steps). The model had learned to keep whichever talker was closer. S4 places the
caller near, mid or far independently of the background talkers and adds real AMI room tone. S3 and S4 keep the real
batch at half weight.

\section{Evaluation}

\textbf{Krisp benchmark}~\cite{krisp2025bench}: 265 real recordings of a main talker with a second voice in the room
(work, call center, phone calls), with transcripts and speaker segment times. Krisp publishes the outputs of its VI
2.5 models, which I score as released. \textbf{AMI}: 134 windows from five meetings and seven people, with the
headset wearer as the caller: ES2003a--d from AMI's training partition and IS1008a from its development partition,
none of them used in my training. I looked at both sets during development: AMI picked among S2 checkpoints, and
Krisp exposed the problems S3 and S4 fix. So neither is blind. \textbf{Blind AMI}: after the model was final, I ran
the same script on the 19 AMI meetings REAL-T uses for evaluation. Eight failed to download, so the 400 windows (16
people) come from 11 meetings, all in AMI's standard test partition. I froze them by hash before running any model.
Each AMI window starts at a 3--15\,s turn of the wearer and runs 4\,s past it while others talk, so the caller always
speaks first, which suits a sample taken from the start.

Transcripts come from Parakeet TDT 0.6B~\cite{xu2023tdt} (v2 on Krisp with Krisp's published text normalization, v3
on AMI). I report three metrics: WER on the caller's words; \emph{missed}, $(S+D)/N$, the share of the caller's words
the transcript loses; and \emph{leak}, the number of recognized words whose midpoint falls where only background
speech is present (more than 0.25\,s from caller speech), per minute of such audio. Intervals are 95\% bootstrap
intervals over people (5,000 draws), uncorrected for multiple comparisons. People in the same meeting are not
independent, so the AMI intervals are, if anything, too narrow.

I compare against Krisp VI 2.5 (four models), ai-coustics Voice Focus 2.2 L and S (local SDK), Sanas Voice
Isolation SE 1.0 (cloud SDK, fed in real time), Hush~\cite{hush2026}, DeepFilterNet3~\cite{schroeter2023dfn3},
TIGER~\cite{xu2025tiger} (in 10\,s chunks, keeping the output whose ECAPA embedding is closest to the first 3\,s of
speech), SAM Audio small~\cite{shi2025samaudio} (a text prompt plus the first 3\,s of speech as an anchor), and the
REAL-TSE checkpoint given the same voice sample as VoxLock. Commercial versions are those available in October 2026.
VoxLock rows are the float32 model on the whole file unless stated. The commercial filters run causally, so
Table~\ref{tab:krisp} also gives \texttt{live()}, which does too.

\section{Results}

\begin{table}[t]
  \caption{Krisp benchmark, 265 real recordings. Missed: caller words lost (\%). Leak: background words per minute
  of background-only speech (85.3\,min). $\Delta$: system minus VoxLock (whole file), 95\% interval; * VoxLock
  significantly better. $^\dagger$264 files.}
  \label{tab:krisp}
  \centering
  \footnotesize
  \setlength{\tabcolsep}{3pt}
  \resizebox{\columnwidth}{!}{%
  \begin{tabular}{@{}lrrrl@{}}
    \toprule
    System & WER & Missed & Leak & $\Delta$WER \\
    \midrule
    VoxLock (open) & \textbf{4.95} & 4.22 & \textbf{0.11} & -- \\
    VoxLock \texttt{live()} & 5.15 & 4.41 & 0.12 & +0.21 [+0.02, +0.43]* \\
    Krisp VI 2.5 Default & 5.30 & 4.21 & 1.52 & +0.35 [$-$0.18, +0.94] \\
    Krisp VI 2.5 Balanced$^\dagger$ & 5.32 & 4.06 & 2.02 & +0.37 [$-$0.19, +0.91] \\
    Krisp VI 2.5 HD & 5.79 & 4.37 & 2.74 & +0.85 [+0.01, +1.81]* \\
    Krisp VI 2.5 Lite & 6.21 & 4.26 & 4.52 & +1.26 [+0.33, +2.07]* \\
    ai-coustics VF 2.2 L & 6.84 & 6.03 & 0.66 & +1.89 [+0.97, +3.10]* \\
    ai-coustics VF 2.2 S & 6.91 & 6.01 & 1.00 & +1.96 [+1.14, +3.03]* \\
    Sanas SE 1.0 & 9.38 & 4.90 & 13.76 & +4.43 [+2.89, +5.70]* \\
    REAL-TSE causal (start) & 9.81 & 5.84 & 13.73 & +4.86 [+2.95, +7.63]* \\
    Hush & 12.09 & 6.27 & 19.58 & +7.14 [+5.36, +9.72]* \\
    DeepFilterNet3 & 13.30 & 6.46 & 22.77 & +8.35 [+6.11, +10.52]* \\
    TIGER + caller pick & 14.72 & \textbf{3.96} & 37.24 & +9.77 [+7.77, +11.75]* \\
    SAM Audio small & 33.77 & 27.98 & 21.15 & +28.83 [+22.90, +34.81]* \\
    No processing & 23.21 & 4.28 & 66.48 & +18.26 [+15.50, +20.82]* \\
    \bottomrule
  \end{tabular}}
\end{table}

\textbf{Krisp benchmark} (Table~\ref{tab:krisp}). VoxLock and Krisp Default are not significantly different on WER,
missed words or leak. Krisp Balanced is not significantly different on WER or missed words but leaks significantly
more. Against Krisp HD the WER difference is borderline, with an interval starting at +0.01. VoxLock has
significantly lower WER than Krisp Lite, both ai-coustics models, Sanas and every open model. Run causally with
\texttt{live()}, VoxLock scores 5.15\%, still below Krisp Default's 5.30\%. In counts, 9 background words reached
the VoxLock transcript in 85.3 minutes of background-only speech, against 130 for Krisp Default and 5,674
unprocessed.

There is a cost when there is nothing to remove. In the phone-call scenario, which has almost no background speech,
VoxLock raises WER from 3.82\% to 5.36\% (Krisp Default: 4.15\%), while in the work and call-center scenarios it
lowers WER from 29.31\% and 25.13\% to 4.83\% and 4.90\%. On 200 windows where only the caller talks, from the 19
blind-set meetings (20 people), VoxLock's WER is 0.31 points above unprocessed audio (7.49\% against 7.18\%;
interval [$-$0.32, +1.33], not significant).

\begin{table}[t]
  \caption{AMI, 134 windows (7 people). Leak counts in parentheses (7.5\,min of background-only speech).
  $\Delta$: system minus VoxLock; * VoxLock significantly better.}
  \label{tab:ami}
  \centering
  \footnotesize
  \setlength{\tabcolsep}{3pt}
  \resizebox{\columnwidth}{!}{%
  \begin{tabular}{@{}lrrrl@{}}
    \toprule
    System & WER & Missed & Leak & $\Delta$WER \\
    \midrule
    VoxLock (open) & \textbf{10.12} & \textbf{6.77} & 2.12 (16) & -- \\
    ai-coustics VF 2.2 S & 11.85 & 8.67 & \textbf{1.06} (8) & +1.73 [+0.09, +3.86]* \\
    ai-coustics VF 2.2 L & 12.13 & 8.88 & 1.72 (13) & +2.00 [$-$0.11, +4.79] \\
    Sanas SE 1.0 & 13.30 & 9.09 & 5.30 (40) & +3.18 [+1.61, +6.71]* \\
    TIGER + caller pick & 22.84 & 9.74 & 36.85 (278) & +12.72 [+9.41, +14.98]* \\
    DeepFilterNet3 & 26.78 & 13.10 & 38.84 (293) & +16.66 [+10.93, +29.50]* \\
    Hush & 33.00 & 11.89 & 68.93 (520) & +22.87 [+15.96, +28.93]* \\
    No processing & 33.03 & 8.05 & 78.34 (591) & +22.91 [+18.54, +34.46]* \\
    \bottomrule
  \end{tabular}}
\end{table}

\textbf{AMI} (Table~\ref{tab:ami}). VoxLock has the lowest WER and the fewest missed words. ai-coustics Voice Focus
2.2 S lets fewer background words through (8 against 16), and that difference is significant. With seven people the
intervals are wide.

\begin{table}[t]
  \caption{Blind AMI set, 400 windows (16 people), frozen before scoring. Leak per minute of background-only speech
  (23.1\,min). Rows are VoxLock unless named. $\Delta$: row minus the first row; * interval excludes zero.
  $^\ddagger$399 windows.}
  \label{tab:blind}
  \centering
  \footnotesize
  \setlength{\tabcolsep}{3pt}
  \resizebox{\columnwidth}{!}{%
  \begin{tabular}{@{}lrrrl@{}}
    \toprule
    System & WER & Missed & Leak & $\Delta$WER \\
    \midrule
    Loudest 3\,s (default) & 10.63 & 8.26 & 1.17 & -- \\
    First 3\,s of speech & 10.22 & 7.79 & 1.13 & $-$0.41 [$-$1.54, +0.09] \\
    \texttt{live()} & 11.97 & 8.67 & 1.78 & +1.34 [$-$0.27, +2.41] \\
    REAL-TSE (start) & 14.75 & 9.08 & 12.18 & +4.12 [+2.49, +5.50]* \\
    No processing & 42.95 & 9.79 & 98.19 & +32.32 [+25.11, +38.88]* \\
    \midrule
    Wrong person's sample$^\ddagger$ & 23.23 & 16.97 & 14.55 & +12.62 [+3.59, +19.61]* \\
    REAL-TSE, wrong sample$^\ddagger$ & 69.92 & 61.26 & 79.68 & +59.31 [+46.20, +66.23]* \\
    \bottomrule
  \end{tabular}}
\end{table}

\textbf{Blind AMI} (Table~\ref{tab:blind}). On windows nobody looked at during development, VoxLock cuts WER from
42.95\% to 10.63\% and leaked words from 98 to 1.2 per minute. Taking the sample from the first 3\,s of speech,
found with Silero VAD~\cite{silero2024vad}, instead of the loudest 3\,s makes no significant difference here; on
Krisp it raises WER by 0.13 points [+0.01, +0.28]. The live path passes each window through unprocessed until it has
heard 3\,s of speech. Other systems are missing from this set: I no longer have the ai-coustics key used for
Table~\ref{tab:ami}, Krisp publishes outputs only for its own benchmark, and I did not run Sanas or the open
baselines on it.

\textbf{Does it use the voice sample?} When the voice sample is a 3\,s turn of another person from the same meeting,
the REAL-TSE checkpoint follows that person and loses 61\% of the caller's words. VoxLock loses 17\% (8\% with the
right sample). So it does use the sample, but it also learned to keep the dominant near-field talker, and a wrong
sample degrades it instead of flipping it to the other person.

\begin{table}[t]
  \caption{Stages. WER and leak on Krisp and the blind AMI set. The two control rows each continue S1 for 1,000
  steps at learning rate $10^{-4}$. $^\S$The Table~\ref{tab:krisp} checkpoint, extracted again for this run; the two
  extractions are not identical, and 40 of 265 files have different error counts.}
  \label{tab:stages}
  \centering
  \footnotesize
  \setlength{\tabcolsep}{4pt}
  \resizebox{\columnwidth}{!}{%
  \begin{tabular}{@{}lrrrrr@{}}
    \toprule
    & & \multicolumn{2}{c}{Krisp} & \multicolumn{2}{c}{Blind AMI} \\
    Stage & Steps & WER & Leak & WER & Leak \\
    \midrule
    REAL-TSE (start) & -- & 9.99$^\S$ & 13.58 & 14.75 & 12.18 \\
    S1 simulated & 4,000 & 5.61 & 2.25 & 11.07 & 1.56 \\
    S2 real crosstalk & 4,000 & 5.76 & 0.20 & 10.82 & 1.60 \\
    S3 long calls & 800 & 5.20 & 0.43 & 10.48 & 1.60 \\
    S4 distance, room & 400 & \textbf{4.95} & \textbf{0.11} & 10.63 & \textbf{1.17} \\
    \midrule
    S1 + 1,000 simulated & & 5.60 & 2.55 & 10.83 & 1.69 \\
    S1 + 1,000 real crosstalk & & 6.06 & 5.35 & 11.28 & 2.47 \\
    \bottomrule
  \end{tabular}}
\end{table}

\textbf{Stages} (Table~\ref{tab:stages}). Most of the gain comes from S1: Krisp WER drops from 9.99\% to 5.61\% and
blind AMI from 14.75\% to 11.07\%. After S1, blind AMI WER moves by at most 0.6 points, and S1, S2 and S3 are not
significantly different from VoxLock there. Blind AMI windows are 7--19\,s of headset audio, so the long-call fade
and the distance shortcut that S3 and S4 fix rarely come up in them. The later stages matter on Krisp, where leak
falls from 2.25 to 0.11 words per minute and WER from 5.61\% to 4.95\% (both significant). I cannot credit real
crosstalk alone for this: the first 1,000 steps of S2 were no better than 1,000 more simulated steps, and leaked
significantly more than S1 on Krisp. The leak drop came later in S2, with the crosstalk-only examples and the
learning-rate decay. This is one training run in which each stage was designed after looking at the previous one,
not a controlled ablation.

\section{Deployment}
\label{sec:deploy}

VoxLock 1.2.1 runs the model as two ONNX graphs. The enroll graph runs once per call on the voice sample and returns
its normalized spectrum and, for every band, the keys and values of the attention over the sample's features. The
chunk graph runs on each chunk of the call with those tensors and the recurrent state, so per chunk it does only
work that depends on the call. In float32, streaming output equals a whole-file pass up to float rounding.

The 8-bit model (recurrent layers quantized) scores 5.11\% WER on Krisp. I measured CPU load on a 60\,s call as
process CPU time over audio time, Python loop included, with one onnxruntime 1.31.1 thread on a Google Cloud
g2-standard-8 (Intel Xeon, 2.2\,GHz). In 80\,ms chunks the call uses 61\% of the thread, in 40\,ms chunks 77\%, and
in 20\,ms chunks 110\%, more than one thread has; on two threads, 20\,ms chunks run at 0.86 times real time (wall
clock). These figures are for steady streaming and leave out the switch described below. Delay is the chunk length
plus the 24\,ms algorithmic latency. On an NVIDIA L4 the float32 model processes a recording 40 times faster than
real time.

At the switch from pass-through to isolation, \texttt{live()} first runs the network over the call so far (up to
30\,s) and drops that output, so the recurrent state has heard the call from its start instead of beginning
mid-sentence. This costs 2--5\,s of compute on one CPU thread, once per call. It runs inside the \texttt{push()} that
completes the voice sample, so on one thread the output stalls for that long at the switch and then lags until it has caught
up. On Krisp, \texttt{live()} scores 5.15\% WER against 4.95\% for the same model on the whole file.

\section{Limitations}
\label{sec:limits}
The model is English-trained, 16\,kHz and single-caller. The automatic voice sample assumes the caller is the
loudest early talker; if a background talker dominates the first seconds, VoxLock keeps the wrong person. Both
comparison sets were seen during development, and every AMI window starts with the caller, which suits the automatic
sample. The Krisp benchmark comes from a vendor whose systems I compare against; its license (CC BY-NC 4.0) rules out
commercial use. All transcripts come from one recognizer family, and all results from one training run.

\section{Conclusion}
A public TSE baseline, fine-tuned on one GPU and given a voice sample from the call, removes background voices for a
recognizer about as well as Krisp's best filters and with lower WER than every other system I tested, holds up on a
blind set of meetings, and runs live on one CPU thread.
\ifdefined\ANON
The model, code and per-file scores are public; links are withheld for review.
\else
The model, ONNX graphs, Python package and per-file evaluation scores are at
\url{https://github.com/arbazsiddiqui/voxlock} and \url{https://huggingface.co/arbazsiddiqui/voxlock}.
\fi

\bibliographystyle{IEEEtran}
\bibliography{refs}

\end{document}